\documentclass[superscriptaddress,twocolumn,prl,preprintnumbers,amsmath,amssymb]{revtex4}
\usepackage{graphicx,afterpage}
\usepackage{hyperref}
\begin{document}

\title{Spin susceptibility of charge ordered  YBa$_2$Cu$_3$O$_y$ across the upper critical field}

\author{R. Zhou}
\affiliation{Univ. Grenoble Alpes, INSA Toulouse, Univ. Toulouse Paul Sabatier, CNRS, LNCMI, 38000 Grenoble, France}
\author{M. Hirata}
\altaffiliation[Present address: ]{Institute for Materials Research (IMR), Tohoku University, Katahira 2-1-1, Aoba-ku, Sendai 980-8577, Japan}
\affiliation{Univ. Grenoble Alpes, INSA Toulouse, Univ. Toulouse Paul Sabatier, CNRS, LNCMI, 38000 Grenoble, France}
\author{T. Wu}
\altaffiliation[Present address: ]{Hefei National Laboratory for Physical Sciences at the Microscale, University of Science and Technology of China (USTC), Anhui, Hefei 230026, P. R. China}
\affiliation{Univ. Grenoble Alpes, INSA Toulouse, Univ. Toulouse Paul Sabatier, CNRS, LNCMI, 38000 Grenoble, France}
\author{I. Vinograd}
\affiliation{Univ. Grenoble Alpes, INSA Toulouse, Univ. Toulouse Paul Sabatier, CNRS, LNCMI, 38000 Grenoble, France}
\author{H.~Mayaffre}
\affiliation{Univ. Grenoble Alpes, INSA Toulouse, Univ. Toulouse Paul Sabatier, CNRS, LNCMI, 38000 Grenoble, France}
\author{S.~Kr\"amer}
\affiliation{Univ. Grenoble Alpes, INSA Toulouse, Univ. Toulouse Paul Sabatier, CNRS, LNCMI, 38000 Grenoble, France}
\author{A.P.~Reyes}
\affiliation{National High Magnetic Field Laboratory, Florida State University, Tallahassee, Florida 32310, USA}
\author{P.L.~Kuhns}
\affiliation{National High Magnetic Field Laboratory, Florida State University, Tallahassee, Florida 32310, USA}
\author{R.~Liang}
\affiliation{Department of Physics and Astronomy, University of British Columbia, Vancouver, BC, Canada, V6T~1Z1}
\affiliation{Canadian Institute for Advanced Research, Toronto, Canada, M5G 1Z8}
\author{W.N.~Hardy}
\affiliation{Department of Physics and Astronomy, University of British Columbia, Vancouver, BC, Canada, V6T~1Z1}
\affiliation{Canadian Institute for Advanced Research, Toronto, Canada, M5G 1Z8}
\author{D.A.~Bonn}
\affiliation{Department of Physics and Astronomy, University of British Columbia, Vancouver, BC, Canada, V6T~1Z1}
\affiliation{Canadian Institute for Advanced Research, Toronto, Canada, M5G 1Z8}
\author{M.-H. Julien}
\email{marc-henri.julien@lncmi.cnrs.fr}
\affiliation{Univ. Grenoble Alpes, INSA Toulouse, Univ. Toulouse Paul Sabatier, CNRS, LNCMI, 38000 Grenoble, France}
\date{\today}

\begin{abstract}
The value of the upper critical field $H_{c2}$, a fundamental characteristic of the superconducting state, has been subject to strong controversy in high-$T_c$ copper-oxides. Since the issue has been tackled almost exclusively by macroscopic techniques so far, there is a clear need for local-probe measurements. Here, we use $^{17}$O NMR to measure the spin susceptibility $\chi_{\rm spin}$ of the CuO$_2$ planes at low temperature in charge ordered YBa$_2$Cu$_3$O$_y$. We find that $\chi_{\rm spin}$ increases (most likely linearly) with magnetic field $H$ and saturates above field values ranging from 20 to 40~T. This result is consistent with $H_{c2}$ values claimed by G. Grissonnanche {\it et al.} [Nat. Commun. {\bf 5}, 3280 (2014)] and with the interpretation that the charge-density-wave (CDW) reduces $H_{c2}$ in underdoped YBa$_2$Cu$_3$O$_y$. Furthermore, the absence of marked deviation in $\chi_{\rm spin}(H)$ at the onset of long-range CDW order indicates that this $H_{c2}$ reduction and the Fermi-surface reconstruction are primarily rooted in the short-range CDW order already present in zero field, not in the field-induced long-range CDW order. Above $H_{c2}$, the relatively low values of $\chi_{\rm spin}$ at $T=2$~K show that the pseudogap is a ground-state property, independent of the superconducting gap.

\end{abstract}

\maketitle

Divergent interpretations of experimental data in the various cuprate families lead to contradictory conclusions as to whether $H_{c2}$ decreases~(\cite{Tafti14} and refs. therein) or increases~(\cite{Yu16} and refs. therein) as hole doping $p$ is reduced below optimal doping. The question is all the more important as it touches on crucial aspects of the cuprate problem: what controls the superconducting transition temperature $T_c$ as a function of doping, what is the microscopic nature of the pseudogap phase, what is the pairing mechanism?

The dispute has become particularly acute in the context of the competition between superconductivity and charge density wave (CDW) order recently revealed by nuclear magnetic resonance (NMR)~\cite{Wu11,Wu13,Wu15}, x-ray scattering~\cite{Ghiringhelli12,Chang12,Achkar12,Hucker14,Blanco14,Gerber15,Chang16,Jang16} and sound velocity~\cite{LeBoeuf13} experiments in underdoped YBa$_2$Cu$_3$O$_y$. An interpretation of thermal conductivity experiments in high magnetic field $H$ argues that $H_{c2}$ becomes as low as $\sim$24~T for $p\simeq0.11$ doping~\cite{Grissonnanche14} (here $p$ changes upon varying the oxygen content $y$ and the relation between $p$ and $y$ is discussed in the Materials \& methods section). This has been argued to be in agreement with a Gaussian-fluctuation analysis of paraconductivity data~\cite{Chang12b,Ando02,Rullier11}, with the value of the vortex melting field extrapolated to $T=0$~\cite{Ramshaw12} and with the observation of the Wiedemann-Franz law~\cite{Grissonnanche16}. According to this interpretation, $H_{c2}$ decreases with decreasing $p$ below optimal doping, and is non-monotonic with a minimum near $p=0.12$ doping. This has been interpreted as suppression of $H_{c2}$ in relation with CDW ordering in the doping range $p\simeq 0.08 - 0.18$, or as the existence of two quantum critical points~\cite{Grissonnanche14}. A differing viewpoint disagrees on such low $H_{c2}$ values and instead argues that superconductivity persists up to at least 45~T, on the basis of thermal conductivity, bulk magnetization and specific heat~\cite{Yu16,Riggs11} measurements.

\begin{figure}[t!]%%%%%%%%%%%%%%%%%%%%%%%%%%%%%%%%%%%%%%%%%%%%%
\centerline{\includegraphics[width=3.4in]{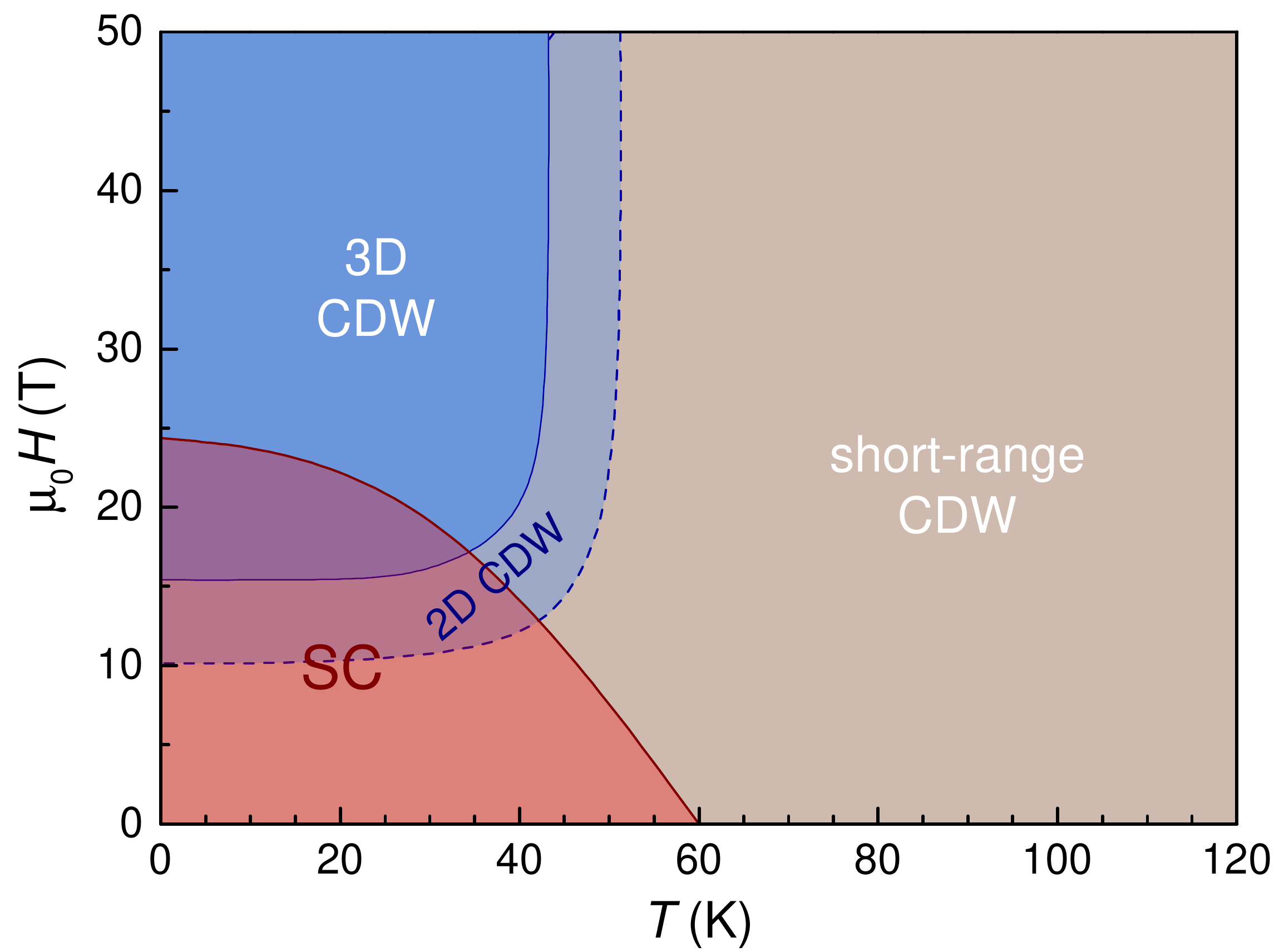}} %%%%%%%%%%%%%%%
 \caption{Schematic field-temperature phase diagram of charge ordered YBCO with $T_c\simeq 60$~K ($p\simeq 0.11$). The boundary between short-range bi-dimensional (2D) correlations and long-range planar correlations (2D CDW) is according to refs.~\cite{Wu11,Wu13,Chang16} and the phase transition toward full three-dimensional long-range order (3D CDW) is according to refs.~\cite{Gerber15,Chang16,Jang16,LeBoeuf13}. The superconducting (SC) transition is according to ref.~\cite{Grissonnanche14}. The experiments reported in this paper follow a vertical line at $T\simeq 2$~K, thus crossing the different CDW phases. }
 \label{phasediag}
\end{figure}

\begin{figure*}[t!]%%%%%%%%%%%%%%%%%%%%%%%%%%%%%%%%%%%%%%%%%%%%%
\centerline{\includegraphics[width=6.9in]{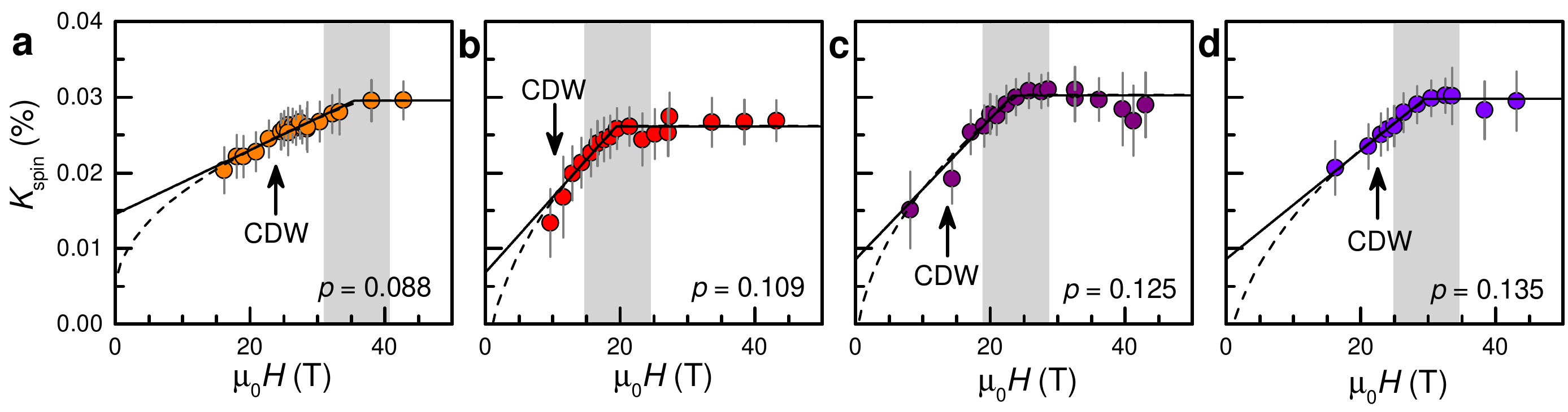}} %%%%%%%%%%%%%%%
 \caption{Magnetic field dependence of (the spin part of) $^{17}$O(3) Knight shift in four different samples at $T=2$~K (a,c,d) and 3~K (b). The grey bars locate the saturation field and the related uncertainty. The arrows mark the CDW onset field of 2D long-range CDW order as determined by NMR, typically 5 to 10~T below the onset of 3D order (see Fig.~1). Continuous and dashed lines correspond to fits with linear ($\alpha=1$) and square-root ($\alpha=1/2$) field dependence, respectively (see text and Eq.~2). Data for O(2) sites, that are less accurate but more sensitive to the CDW, are entirely consistent with O(3) data (Fig. S3).}
\label{shift}
\end{figure*}
%%%%%%%%

The determination of $H_{c2}$ is actually complicated by the absence of sharp experimental signature of the transition to the normal state in high fields and by the limitations of each experimental probe. For instance, macroscopic measurements can be biased by doping inhomogeneity (especially electrical transport that is only sensitive to the lowest-resistance path in the sample) or, in the case of YBa$_2$Cu$_3$O$_y$, by magnetism from the chains. Furthermore, additional complication potentially arises from the presence of the field-induced phase transition towards long-range CDW ordering in the phase diagram of YBa$_2$Cu$_3$O$_y$ (Fig.~\ref{phasediag}).

Here, we provide a new perspective on this question by using a microscopic, site-selective probe in underdoped YBa$_2$Cu$_3$O$_y$ single crystals (information about the samples and the experiments can be found in the 'Materials \& methods' section and in the Supporting information). Specifically, we deduce the field dependence of the static, uniform spin susceptibility $\chi_{\rm spin} = \chi_{\rm spin}(q=0,\omega=0)$ from NMR measurements of the spin part of the Knight shift $K_{\rm spin}$ at planar $^{17}$O sites (see supporting information for details concerning the subtraction of the orbital shift from the total shift data). These two quantities are directly proportional through:
\begin{equation}
K_{\rm spin} = \frac{A}{g\mu_B} \chi_{\rm spin}
\end{equation}
where $A$ is the hyperfine coupling constant and $g$ the Land\'e factor (all quantities are tensors that are diagonal in the frame of the crystallographic axes, but indices have been dropped for the sake of simplicity).

In a simple metal, $\chi_{\rm spin}$ is proportional to $N(E_F)$, the density-of-states (DOS) at the Fermi level $E_F$. In the underdoped cuprates, the situation is much more complex as $\chi_{\rm spin}$ has a relatively weak and smooth evolution from the antiferromagnetic insulator (where it reflects the response of localized Cu$^{2+}$ moments) to the overdoped metallic phase (where it is more likely to be proportional to $N(E_F)$)~\cite{Johnston89,Alloul89}. Nevertheless, at low temperature ($T$), $\chi_{\rm spin}$ should be dominated by $N(E_F)$, for the following reasons.

In a superconductor with singlet pairing, $\chi_{\rm spin}$ is considered to vanish at $T=0$. This, however, is no longer true in the mixed state, under an applied magnetic field. In $d$-wave superconductors, the Doppler shift of nodal quasiparticles (i.e. those with momenta close to the position of the gap nodes) outside vortex cores is responsible for a residual $N(E_F)$ that increases with magnetic field as $\sqrt{H}$~\cite{Volovik,Moler94,Hussey02}. In addition to this, a field-independent residual $N(E_F)$ can arise from disorder~\cite{Ishida91}. Experimentally, the field dependence of $K_{\rm spin}$ in several cuprates is consistent with a $\sqrt{H}$ dependence~\cite{Zheng99,Zheng02,Kawasaki10}. Therefore, the field dependence of $\chi_{\rm spin}$ at low $T$ should reflect the change in $N(E_F)$, even though $\chi_{\rm spin}$ is, in general, not related to $N(E_F)$ in a simple way.

\begin{figure}[b!]%%%%%%%%%%%%%%%%%%%%%%%%%%%%%%%%%%%%%%%%%%%%%
\centerline{\includegraphics[width=3in]{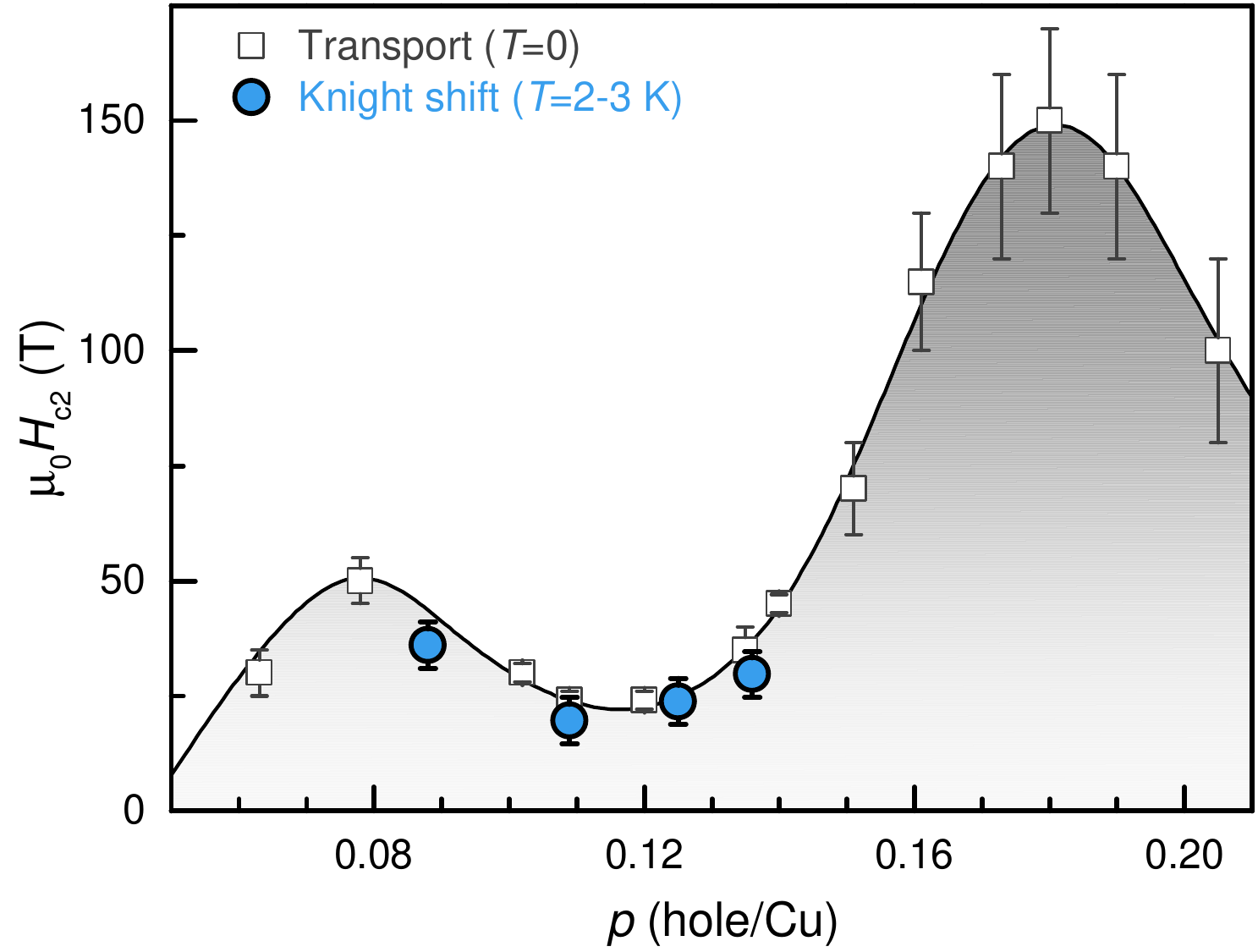}} %%%%%%%%%%%%%%%
 \caption{Saturation field in Knight shift measurements (blue dots, this work) at $T=2$~K (3~K for the $p=0.109$ sample), compared to $H_{c2}$ values extrapolated at $T=0$ from resistivity data~\cite{Ramshaw12,Grissonnanche14}. The remarkable agreement leads to the identification of the saturation field as $H_{c2}$. The solid line is a guide to the eye.}
 \label{hc2diag}
\end{figure}

%%%%%%%%%%%%%%%

\section*{Upper critical field}

We have measured the total Knight shift in four different crystals and have determined its spin part $K_{\rm spin}\propto \chi_{\rm spin}$ by subtracting the orbital contribution, while the contribution from diamagnetic shielding was found to be negligible at the fields used - see Materials and Methods (MM) section and Supporting Information (SI) for details. Figure~\ref{shift} summarizes the central result of this work: after initially increasing upon increasing field, we observe that $\chi_{\rm spin}$ saturates above a certain field in all the samples. Since $\chi_{\rm spin}$ is expected to increase with $H$ as long as superconductivity is present, the most natural explanation for the saturation is that the upper critical field $H_{c2}$ has been reached. A similar observation has previously been made in optimally doped Bi2201~\cite{Mei12}.

In order to precisely determine the saturation field $H_{\rm sat}$, we perform global fits whereby $K_{\rm spin}(H)$ data are fitted to a $H^\alpha$ dependence below $H_{\rm sat}$ and to a constant value above $H_{\rm sat}$, while treating $H_{\rm sat}$ as a fitting parameter. Namely, we have:
\begin{equation}
\begin{split}
K_{\rm spin} = K_0 + K_H (H/H_{\rm sat})^{\alpha}~\text{for}~H \leq H_{\rm sat}\\
K_{\rm spin} = K_0 + K_H ~\text{for}~H > H_{\rm sat}
\end{split}
\end{equation}
We expect this approximation to provide an accurate, unbiased way to determine $H_{\rm sat}$, even though the crossover between the two field regimes must be rounded in reality. Note that we assume here a single field dependence from $H=0$ up to $H_{\rm sat}$ but since our data do not cover low fields, we cannot exclude a more complex dependence.

As Fig.~\ref{hc2diag} shows, the saturation field turns out to agree with the values of $H_{c2}$ determined from transport measurements~\cite{Grissonnanche14,Ramshaw12}, for each of the four samples and independently of the value of $\alpha$. The slightly lower values from NMR are likely attributed to the fact that our data are taken at 2-3~K while transport data are extrapolated to $T=0$. A recent specific heat study~\cite{Marcenat15} of ortho-II YBa$_2$Cu$_3$O$_{6.54}$ ($p\simeq0.11$) also finds a saturation at fields above $\sim 20$~T at $T=7$~K. Furthermore, the absence of a significant diamagnetic shielding contribution to the apical O(4) Knight shift is consistent with reduced $H_{c2}$ values (see methods and Fig.~S1).

Our measurements are therefore consistent with the idea that $H_{c2}$ in YBa$_2$Cu$_3$O$_{y}$ has a pronounced depression around $p=0.11-0.12$ where it reaches values as low as $\sim$20~T. This is the main result of this work. Even though $H_{c2}$ might be enhanced by the presence of quantum critical points at $p\simeq0.08$ and $p\simeq0.18$ doping~\cite{Sebastian08,Ramshaw15,Grissonnanche14}, comparison with other compounds~\cite{Tafti14} shows that $H_{c2}$ is anomalously low where the CDW is strongest while $T_c$ still around 60~K in YBCO.

As remarked in ref.~\cite{Wu13}, such $H_{c2}$ values are close to the field scale of $\sim$30~T at which the CDW order parameter saturates. The difference between 20 and 30~T at $\simeq 0.11$ might be due to the presence of superconducting fluctuations. These have also been recently proposed~\cite{Yu15} as a possible explanation of the diamagnetic response reported in high fields (\cite{Yu16} and refs. therein).

\begin{figure}[t!]%%%%%%%%%%%%%%%%%%%%%%%%%%%%%%%%%%%%%%%%%%%%%
\centerline{\includegraphics[width=3.4in]{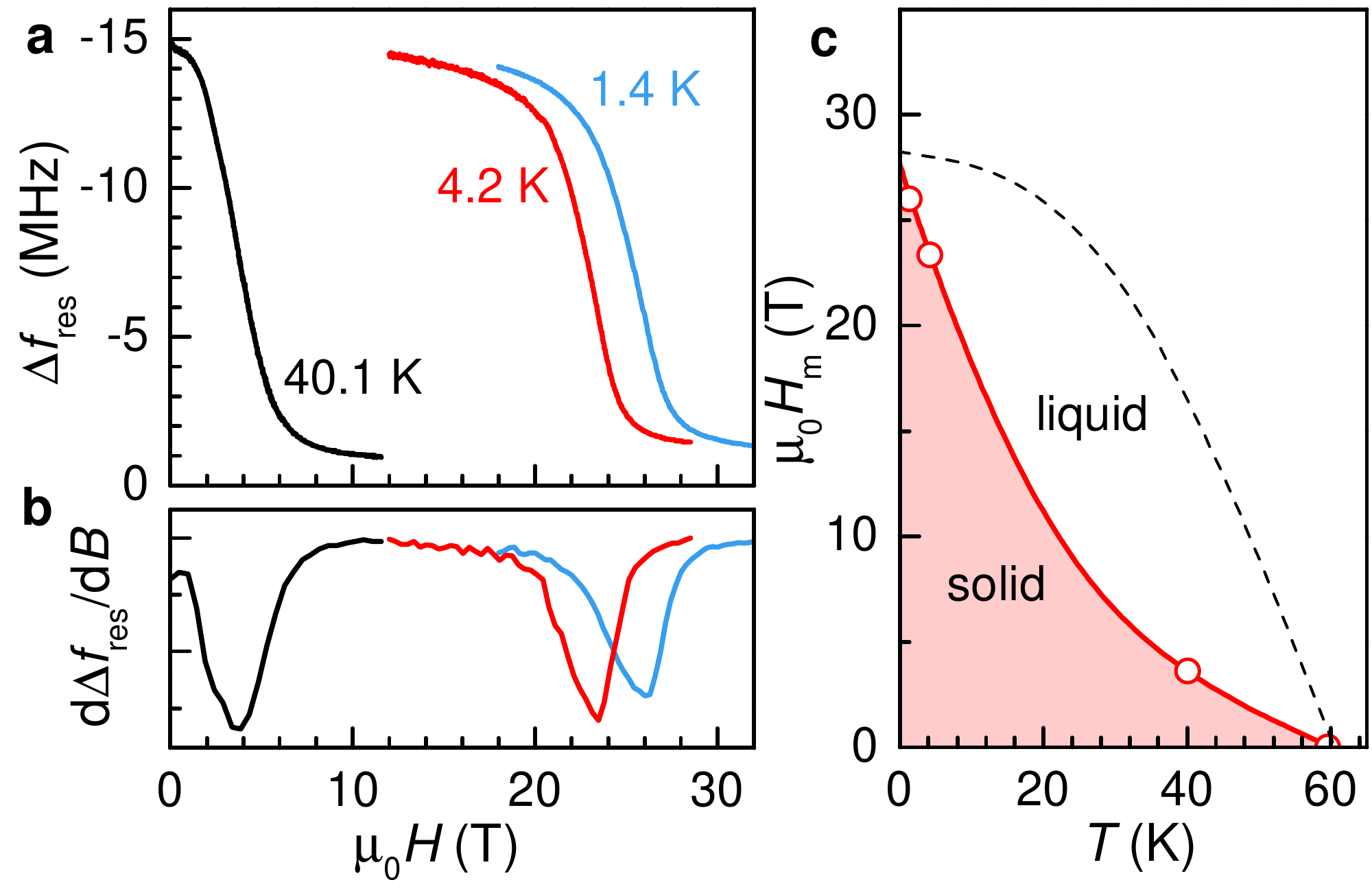}} %%%%%%%%%%%%%%%
 \caption{(a) Field dependence of the change of resonance frequency $\Delta f_{\rm res}$ of the NMR tank circuit (see methods), in YBa$_2$Cu$_3$O$_{6.54}$ ($p=0.104$)~\cite{Wu11}. (b) Derivative of $\Delta f_{\rm res}$. (c) Vortex phase diagram. Open circles correspond to the peak positions in panel (b), which defines the transition between vortex liquid and vortex solid phases. }
 \label{tuning}
\end{figure}

\subsection*{Vortex melting transition}

Part of the controversy about $H_{c2}$ actually lies in a disagreement on the melting transition of the vortex lattice. The $H_{c2}$ values found here appear to be in contradiction with the lower limit $H_{c2}>45$~T inferred from a specific heat experiment~\cite{Riggs11} but they are also in contradiction with the irreversibility observed in the magnetization data of Yu {\it et al.}~\cite{Yu16} (see also~\cite{Sebastian08}). This latter has been taken as evidence of a secondary vortex solid phase extending above 40~T in the $T\rightarrow0$ limit.

\begin{figure*}[t!]%%%%%%%%%%%%%%%%%%%%%%%%%%%%%%%%%%%%%%%%%%%%%
\centerline{\includegraphics[width=7in]{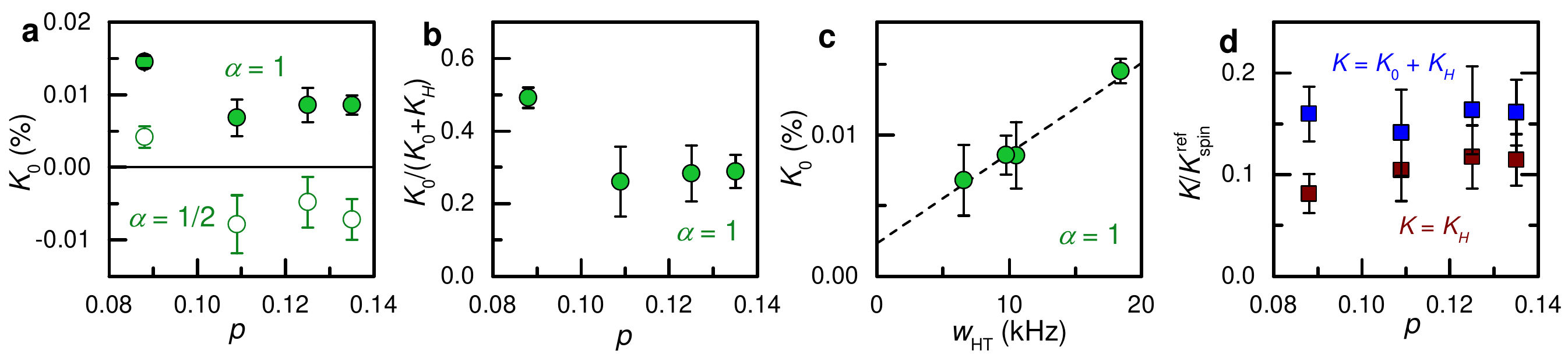}} %%%%%%%%%%%%%%%
 \caption{(a) Residual $K_{\rm spin}$ in the $H=0$ limit, $K_0$, obtained from fits of the data in Fig.~2 to Eq.~2 with $\alpha =1$ (filled circles) and 1/2 (open circles), as a function of doping $p$. (b) $K_0$ relative to $K_0+K_H=K_{\rm spin}(H=H_{c2})$ ($\alpha =1$). (c) Correlation between $K_0$ and the width of the $^{17}$O line at 250~K that quantifies the amount of disorder ($\alpha =1$).  Width data are for the ($m_I=-1/2 \leftrightarrow -3/2$ transition) transition of the O(2EF) sites (see Fig.~1 in ref.~\cite{Wu15}). (d) $K_{\rm spin}(H_{c2}) = K_0+K_H$ and $K_{\rm spin}(H_{c2}) - K_0 = K_H$ relative to the normal state $K_{\rm spin}^{\rm ref}$ in near optimally-doped YBa$_2$Cu$_3$O$_{7-\delta}$~\cite{Horvatic89}.}
 \label{residual}
\end{figure*}

%%%%%%%%%%%%

Our measurements of surface conductivity at $T=1.4$~K (Fig.~\ref{tuning}a) show no evidence of a vortex solid above 30~T: only one jump is seen and the frequency change across the transition is identical to that at 40~K: $\Delta f_{\rm res} \simeq14$~MHz. Furthermore, the absence of loss of NMR signal intensity down to at least 1.4~K in the same sample~\cite{Wu11} also argues against a vortex solid at 1.4~K/30~T.

The irreversibility observed in torque measurements of the bulk magnetization thus remains, to our view, mysterious. We can only conjecture that disorder, which is present even in the best YBCO single crystals~\cite{Wu16}, plays a role. In particular, for two-dimensional systems in the presence of disorder, it has been shown that rare regions with anomalously high $H_{c2}$ values become Josephson-coupled at very low $T$~\cite{Spivak,Gafilms,Lesueur,Popovic}. This gives rises to two stages in the superconducting-to-normal transitions measured by resistivity {\it vs.} field~\cite{Spivak,Gafilms,Lesueur,Popovic}. The lower one is insensitive to Josephson currents and thus characteristic of the cleanest parts of the sample. Unlike bulk measurements, the NMR Knight shift should be sensitive only to this "intrinsic" $H_{c2}$. Indications that inhomogeneity is relevant in YBa$_2$Cu$_3$O$_y$ is found in the absence of narrowing of the melting transition on cooling (Fig.~\ref{tuning}b) and in resistive measurements  {\it vs.} field showing broad superconducting-to-normal transitions as compared to the stoichiometric sibling YBa$_2$Cu$_4$O$_8$~\cite{LeBoeuf07}. Another possibility is that there exist tiny regions with different oxygen content, like the surface of the crystals, grain boundaries or small inclusions of different oxygen order.

%%%%%%%%%%%%%%

\subsection*{Density of states in the superconducting state}

We now discuss the field-dependence of the spin susceptibility below $H_{c2}$ and how it extrapolates to zero field.

NMR studies of cuprate samples~\cite{Zheng02,Kawasaki10} have found that $K \propto \sqrt{H/H_{c2}}$ at $T\ll T_c$, as expected for the Doppler shift of nodal quasiparticles outside vortex cores~\cite{Volovik}. Here in YBa$_2$Cu$_3$O$_y$ single crystals, screening of the radio-frequency severely degrades the signal-to-noise ratio in the superconducting state. This prevents investigation of the low-field range ($\mu_0 H \lesssim 10$~T) where the $\sqrt{H}$ dependence would be most easily distinguishable from a linear dependence. Unsurprisingly, fits to Eq.~2 with $\alpha=1$ and $\alpha=1/2$ (Fig.~2) work equally well. Nevertheless, they lead to somewhat different results regarding $K_0$, a parameter that represents the residual $K_{\rm spin}$ in the zero field and $T\rightarrow 0$ limit (usually not considered in NMR studies because it is impossible to distinguish from the orbital shift $K_{\rm orb}$ in the absence of accurate measurement of the field dependence).

For $\alpha=1/2$ (Fig.~\ref{residual}a), $K_0$ is negative for three of the samples, which is unphysical for $K_{\rm spin}$ (there are neither diamagnetic nor orbital contributions here, see SI). A much larger (positive) value is found for the $y=6.47$ ($p=0.088$) sample which we suspect to be due to an important amount of disorder in this sample. Indeed, we find that $K_0$ scales with the $^{17}$O linewidth at high temperature (Fig.~\ref{residual}c). This latter is always a reliable indication of the level of disorder and is indeed larger in the $p=0.088$ sample than in the other three.

For $\alpha=1$, $K_0$ values are larger than for $\alpha=1/2$ and positive for the four samples (Fig.~\ref{residual}a). We find that the residual $K_0$ at $H=0$ amounts to $\sim30\pm10\%$ of the high-field value $(K_0+K_H$), for $0.109 \leq p \leq 0.135$ (Fig.~\ref{residual}b). This agrees with the residual term in the specific heat of 38\% of the high field value~\cite{Marcenat15}.

The agreement with specific heat data at $p\simeq 0.11$ and the more physical $K_0$ values lead us to favor the $\alpha=1$ scenario, that is, $N(E_F)$ is linear in field, which is also in agreement with recent specific heat data~\cite{Marcenat15}. Nevertheless, we do not claim that our data demonstrate that $\chi_{\rm spin}$ is linear in $H$ or that $K_0$ is finite. Indeed, the $K_0$ values are of the order of uncertainties in $K_{\rm orb}$ values, our data are not strictly at 0~K but at 2-3~K, and we reiterate that a more complex field dependence cannot be excluded (for instance $\propto \sqrt{H}$ below $\sim H_{c2}/2$ and $\propto H$ above). Theoretically, a linear field dependence is expected when the Zeeman energy of the nodal quasiparticles exceeds the Doppler energy (ref.~\cite{Vekhter01} and refs. therein). This could happen when magnetic order sets in, as observed in CeCoIn$_5$~\cite{Koutroulakis08}, or if the Doppler energy is anomalously reduced, as could happen if the field induces changes in the Fermi surface. The latter scenario would be more likely in charge ordered YBCO as we know that there is a field-dependent CDW that reconstructs the Fermi-surface, without any concomitant SDW ordering, even at high fields~\cite{Wu11,Wu13}.

Because oxygen-ordered YBCO has far less disorder than any other cuprate family, the presence of a comparatively large residual DOS in zero field specific heat data has remained unexplained. If $N(E_F)$ is indeed linear in $H$, our results shed new light on this issue. First, NMR being a local probe, that $K_0 \neq 0$ in planar $^{17}$O data would imply that the corresponding residual DOS in zero-field does not originate from the CuO chains. Furthermore, the finite intercept in Fig.~\ref{residual}c would suggest that a fraction of the residual DOS is intrinsic, {\it i.e.} not related to disorder. While this would be consistent with the fact that the residual DOS in YBCO is larger than in, {\it e.g.}, the more disordered La$_{2-x}$Sr$_x$CuO$_4$~\cite{Riggs11}, we caution, however, that the data are also compatible within error bars with a zero intercept.

\subsection*{Connecting with pseudogap and CDW phenomena}

The putative relationship between the pseudogap and the CDW, as well as their respective relationship with superconductivity, has been much discussed in recent years. It is thus interesting to observe how these phenomena affect the spin susceptibility at low temperature, especially upon quenching superconductivity with the field.

A number that may be benchmarked against theoretical predictions is the fraction of DOS left at low $T$ by the pseudogap and by a hypothetical CDW gap, in the absence of superconductivity. For this, $K_{\rm spin}(H=H_{c2})$ should be compared to $K_{\rm spin}$ for a Fermi surface unaffected by the pseudogap. The problem is that $K_{\rm spin}(T=T^*$) is unlikely to be a correct measure of $N(E_F)$ at $T^*$ for underdoped samples and anyway $K_{\rm spin}(T^*)$  cannot be measured here due to oxygen diffusion above room $T$ (defining $T^*$ as the temperature at which $\chi_{\rm spin}$ has a maximum). We thus take as a reference the value of $K_{\rm spin}$ of near-optimally doped YBCO, which is $T$ independent in the normal state~\cite{Horvatic89}. Then, we find that the DOS at $H_{c2}$ and $T=0$ represents a small fraction of this "non-interacting DOS", independently of doping: $K_{\rm spin}({\it H_{c2})}/K_{\rm spin}^{\rm ref}\simeq$16\% (Fig.~\ref{residual}d). This number is independent of the value of $\alpha$, {\it i.e.} of the $H$ dependence below $H_{c2}$. The ratio drops to 10\% if $K_0$ for the $\alpha=1$ scenario is subtracted (Fig.~\ref{residual}d).

\begin{figure}[t!]%%%%%%%%%%%%%%%%%%%%%%%%%%%%%%%%%%%%%%%%%%%%%
\centerline{\includegraphics[width=3.4in]{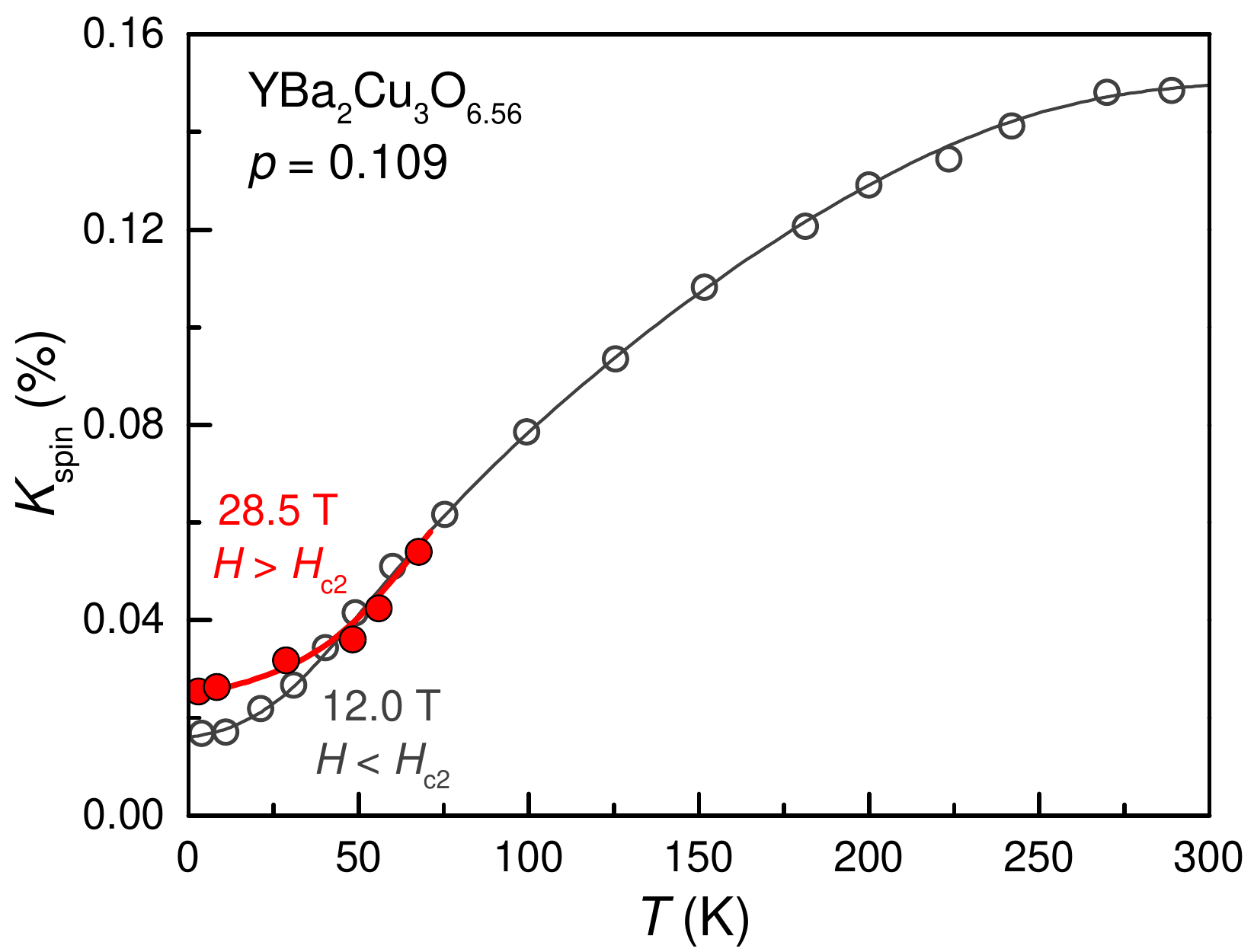}} %%%%%%%%%%%%%%%
 \caption{Temperature dependence of the spin part of the Knight shift $K_{\rm spin}$ at two different fields, below and above $H_{c2}$ for O(3) sites in YBa$_2$Cu$_3$O$_{6.56}$. The data are taken in the same conditions as those in Figs. 2 and 5 (that is, with the field slightly tilted off the $c$ axis) whereas data in ref.~\cite{Wu13} were for $H$ exactly parallel to $c$ and $K_{\rm orb}$ was not subtracted. Lines are guides to the eye.}
 \label{shift_temp}
\end{figure}

In doing these comparisons, we keep in mind that relative DOS values are unlikely to be equal to the ratio of $\chi_{\rm spin}$ values at different doping levels. Even in a Fermi-liquid, $\chi_{\rm spin}$, unlike the specific heat, is not only proportional to $N(E_F)$ but also to correction factors that are correlation-strength, and thus doping, dependent. This may be the reason why the factor-3 enhancement of $m^*$ on decreasing $p$ from $\sim0.10$ to $\sim$0.09~\cite{Sebastian10} is not reflected in a corresponding increase in $\chi_{\rm spin}$ (Fig.~\ref{shift}). Nevertheless, there is no doubt that $\chi_{\rm spin}$ remains small above $H_{c2}$. For instance, we observe that $K_{\rm spin}(H_{c2})/K_{\rm spin}$(285~K) $\simeq$ 20\% (Figs.~\ref{shift_temp} and Fig.~S4).

The small DOS remaining in high fields demonstrates that the non-superconducting ground state has a large pseudogap, separate from the superconducting gap. Similar observation was already made in ref.~\cite{Wu13} for $p=0.109$ but our new data now show that this conclusion is generically robust for charge-ordered YBCO. In order to visualize more directly how small field-induced changes in $\chi_{\rm spin}$ are as compared to those induced by the pseudogap, we show the $T$ dependence of $\chi_{\rm spin}$ at two different fields, below and above $H_{c2}$ (Fig.~\ref{shift_temp}). Interestingly, the remaining DOS fraction of 10 to 20\% at $H_{c2}$ in YBCO is distinctly smaller than in underdoped Bi2201 for which $K_{\rm spin}(H=H_{c2})/K_{\rm spin}(T^*)\simeq$~25 to 50\%~\cite{Kawasaki10}.

The data in Fig.~\ref{shift} (see also supporting information) also show that $K_{\rm spin}$ does not undergo any major change at either $H_{\rm CDW}^{2D}\simeq10$ to 20~T or $H_{\rm CDW}^{3D}$ (5 to 10~T higher than $H_{\rm CDW}^{2D}$). This result parallels the insensitivity of both the specific heat~\cite{Riggs11,Marcenat15} and the coefficient of the $T^2$ term in the in-plane resistivity~\cite{Proust16} to the CDW transition in high fields.

The absence of a decrease of $K_{\rm spin}$ at the CDW transition indicates that long-range CDW order does not produce substantial gapping of quasiparticle excitations. This means either that the pseudogap is totally unrelated to the CDW~\cite{Verret17} or that the short-range CDW modulations, that are present at any field, already deplete $N(E_F)$ before the transition towards long-range order intervenes.

The apparent absence of impact of the CDW transitions on the field dependence of $K_{\rm spin}$ suggests that there is no strong modification of either $H_{c2}$ or $N(E_F)$ at these transitions. Hence, both the drop of $H_{c2}$ and the Fermi-surface reconstruction~\cite{LeBoeuf07,Sebastian15} must be essentially rooted in the presence of short-range (but static) CDW correlations in zero field (which does not exclude that these phenomena be field-dependent as short-range CDW modulations indeed are below $T_c$). That the reduction of $H_{c2}$ does not result from a particular fragility of the superconducting state when it coexists with long-range CDW order is consistent with the known depression of $T_c$ in zero-field~\cite{Liang06} as well as with the observation that similar $H_{c2}$ values, including the drop around $p=0.12$, are deduced from a Gaussian-fluctuation analysis of paraconductivity data in zero field~\cite{Tafti14,Chang12b}. Reconstruction of the Fermi surface by the short-range, rather than by the long-range, CDW order agrees with a number of works~\cite{Chan16,recons1,recons2,recons3,recons4,recons5,recons6,recons7,recons8,Laliberte17}.

\section{Methods}
\subsection*{\bf Samples}
The Knight shift $K$ was measured in four oxygen-ordered, detwinned single crystals: YBa$_2$Cu$_3$O$_{6.47}$ (hole doping $p = 0.088$, ortho-II oxygen order), YBa$_2$Cu$_3$O$_{6.56}$ (hole doping $p = 0.109$, ortho-II oxygen order), YBa$_2$Cu$_3$O$_{6.68}$ (hole doping $p = 0.125$, ortho-VIII oxygen order) and YBa$_2$Cu$_3$O$_{6.77}$ (hole doping $p = 0.135$, ortho-III oxygen order). The hole-doping $p$ is deduced from the relation between $p$ and the superconducting transition temperature $T_c$ established in ref.~\cite{Liang06} for these oxygen-ordered crystals. The crystals were enriched with the $^{17}$O isotope (nuclear spin $I = 5/2$). More information about our samples and experimental methods can be found in refs.~\cite{Liang12,Wu16}.

\subsection*{\bf Experimental conditions}
The field was tilted by an angle $\theta\simeq 16^{\circ}$ off the $c$-axis in order to minimize overlap between O(2) and O(3) lines. Here, the quoted field values correspond to the $c$-axis projection, namely they have been corrected by a factor $\cos(16^{\circ})\simeq0.96$ with respect to nominal values. This $5\pm1$\% correction does not affect any of the conclusions of this work. The field was applied at low temperatures ($T \simeq 2-3$~K), therefore crossing three different regions of the field-temperature phase diagram of YBCO (Fig.~\ref{phasediag}).

\subsection*{\bf Knight shift determination}
$K_{\rm spin}$ values are obtained by subtracting the so-called orbital part $K_{\rm orb}$ and the diamagnetic shielding contribution $K_{\rm dia}$ from the measured total $K$:
\begin{equation}
K = K_{\rm spin} + K_{\rm orb} + K_{\rm dia}
\end{equation}
$K_{\rm dia}$ should be identical at different $^{17}$O positions. Therefore, the absence of any temperature dependence of $K$ at the apical O(4) sites (Fig.~S1) indicates that $K_{\rm dia}$ is negligible at the fields used here. As to the orbital part, we took $^{17}K^{c}_{\rm orb}=-0.014$\% and $^{17}K^{b}_{\rm orb}=0.044$\% from ref.~\cite{Takigawa89} and used the relation $K_{\rm orb}(\theta)=K^c_{\rm orb} \cos^2(\theta)+K^b_{\rm orb} \sin^2(\theta)$. These $K_{\rm orb}$ values are consistent with our measured shift anisotropies (Fig.~S2).

Technical details about Knight shift measurements can be found in Supporting Information and a full $^{17}$O NMR spectrum is shown in ref.~\cite{Zhou17}.

\subsection*{\bf Vortex transition}
Vortex melting can be probed with the NMR setup: the rapid change in resistivity across the melting line modifies the skin depth of the sample, which in turn produces a change of the inductance of the NMR coil surrounding the sample~\cite{PDO}. This then shifts the resonance frequency $f_{\rm res}$ of the NMR tank circuit. In its basic principle, the method is analogous to contactless measurements using diode-oscillators~\cite{PDO}.

\section{Acknowledgments}
We thank J.R.~Cooper, S.~Julian, I.~Kokanovi\'c, T.~Klein, D.~LeBoeuf, M.~Le Tacon, J.~Lesueur, C.~Marcenat, A.~McCollam, V.~Mitrovi\'c, B.~Ramshaw, T.~Senthil, C.~Proust, L.~Taillefer, J.F.~Yu for useful discussions and S.~Kivelson for letting us know ref.~\cite{Spivak}.

Work in Grenoble was supported by the French Agence Nationale de la Recherche (ANR) under reference AF-12-BS04-0012-01 (Superfield), by the Laboratoire d'excellence LANEF in Grenoble (ANR-10-LABX-51-01) and by the Universit\'e J. Fourier - Grenoble (now Universit\'e Grenoble Alpes). Part of this work was performed at the LNCMI, a member of the European Magnetic Field Laboratory (EMFL). A portion of this work was performed at the National High Magnetic Field Laboratory, which is supported by the National Science Foundation Cooperative Agreement No. DMR-1157490, the State of Florida. Work in Vancouver was supported by the Canadian Institute for Advanced Research and the Natural Science and Engineering Research Council.

\bibliography{hc2_pnas}

\clearpage
\newpage

\section{Supporting Information}

\renewcommand\thefigure{S\arabic{figure}}
\setcounter{figure}{0}

\renewcommand\theequation{S\arabic{equation}}
\setcounter{equation}{0}

\begin{figure}[b!]%%%%%%%%%%%%%%%%%%%%%%%%%%%%%%%%%%%%%%%%%%%%%
\centerline{\includegraphics[width=3.in]{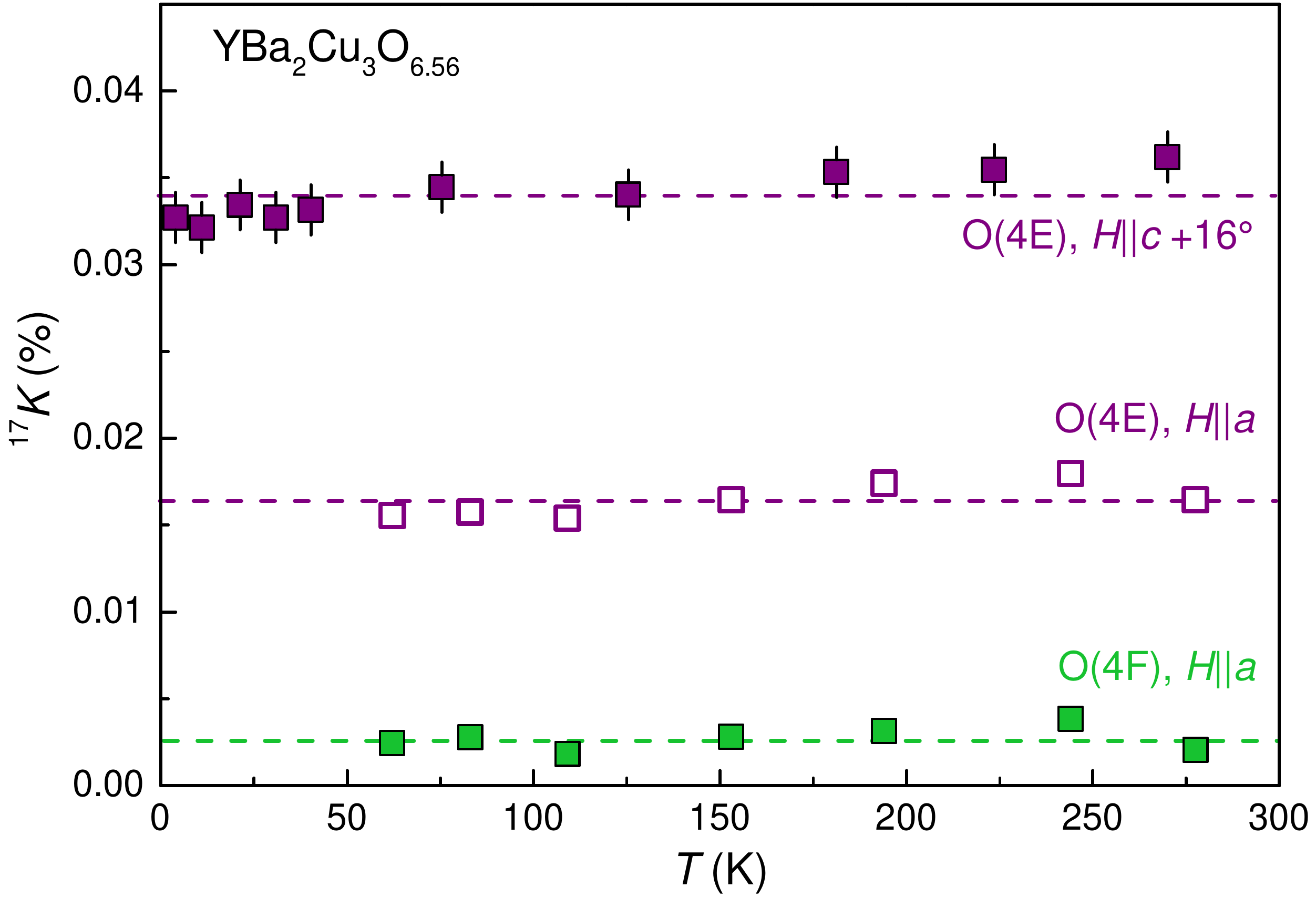}} %%%%%%%%%%%%%%%
 \caption{Total Knight shift at the apical O(4) site as a function of temperature in YBa$_2$Cu$_3$O$_{6.56}$ ($p=0.109$). In the ortho-II structure, there are two inequivalent O(4) sites: O(4E) below empty chains and O(4F) below full chains. O(4F) lines is much broader than O(4E) and so more difficult to measure. Dashed lines are guides to the eye. Data were taken in a field of 12~T.}
 \renewcommand{\figurename}{S1}
 \label{suppl_apical}
\end{figure}

\subsection*{\bf Knight shift measurements}

The central lines of the different sites (O(2), O(3), apical O(4) and chain O(1)) overlap at low temperatures and thus cannot be reliably used for measuring $K$. Therefore, $K$ values were estimated from the position of the broader, satellite transitions (see ref.~\cite{Zhou17} for a typical example of full $^{17}$O spectrum).

\begin{figure}[t!]%%%%%%%%%%%%%%%%%%%%%%%%%%%%%%%%%%%%%%%%%%%%%
\centerline{\includegraphics[width=3in]{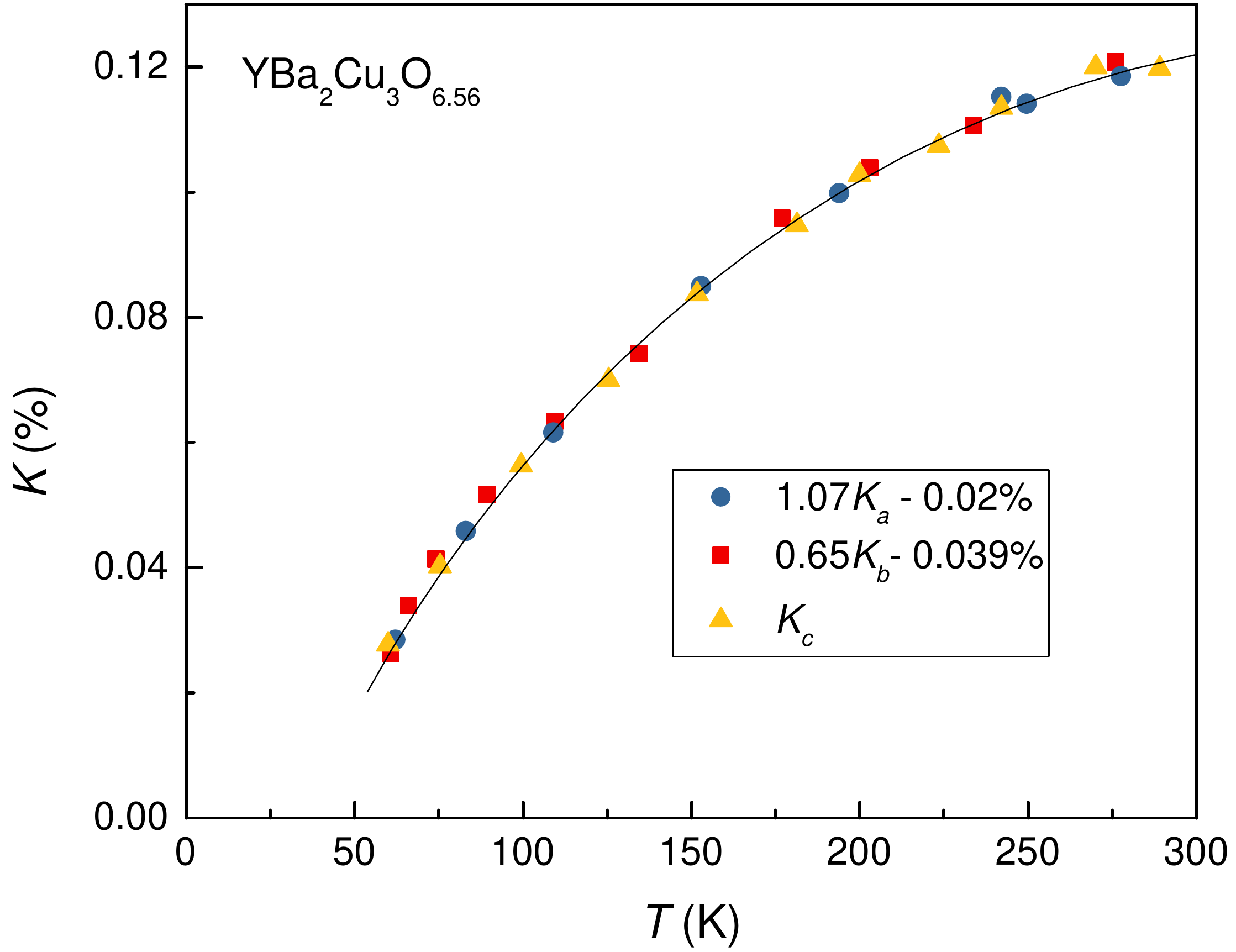}} %%%%%%%%%%%%%%%
 \caption{O(3) total Knight shift data along the three crystallographic axes $a$, $b$ and $c$, appropriately scaled, in YBa$_2$Cu$_3$O$_{6.56}$ ($p=0.109$). The line is a guide to the eye.}
 \label{suppl_scaling}
\end{figure}

\begin{figure*}[t!]%%%%%%%%%%%%%%%%%%%%%%%%%%%%%%%%%%%%%%%%%%%%%
\centerline{\includegraphics[width=6.5in]{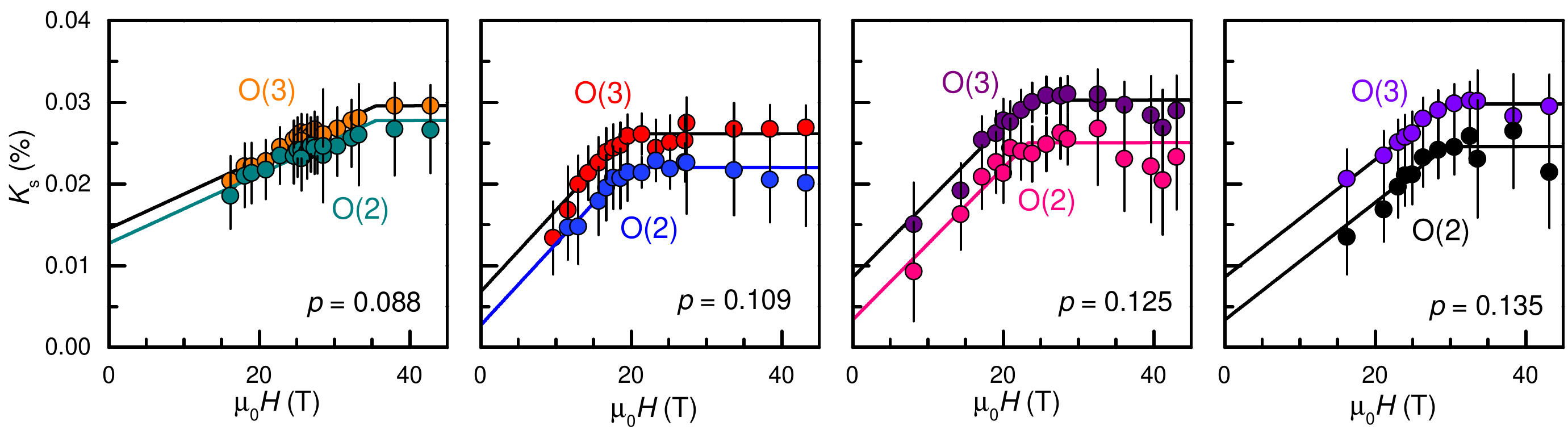}} %%%%%%%%%%%%%%%
 \caption{Comparison of $K_{\rm spin}$ data at O(3) (same data as in Fig.~2) and O(2) sites. Lines are guides to the eye. Both sets of data are essentially identical (same saturation field, same field-dependence, same insensitivity to the CDW transition). $K$ values for O(2) are lower than those for O(3), which we attribute to a systematic error in the fitting, due to the uncertainty generated by the absence of splitting of the low-frequency satellites. }
 \label{suppl_o2o3}
\end{figure*}

\begin{figure}%[b!]%%%%%%%%%%%%%%%%%%%%%%%%%%%%%%%%%%%%%%%%%%%%%
\centerline{\includegraphics[width=3.in]{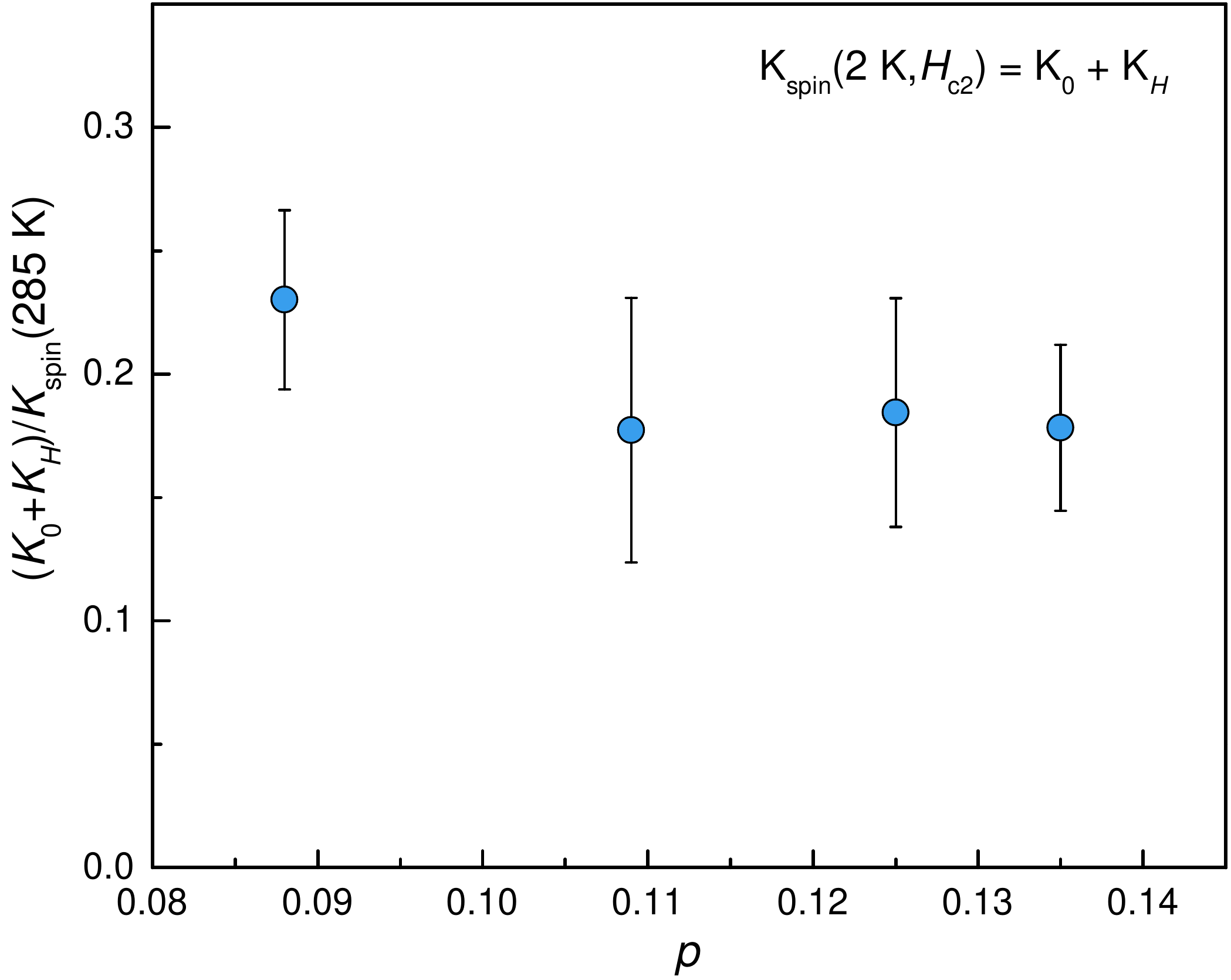}} %%%%%%%%%%%%%%%
 \caption{"Residual" (spin part of the) Knight shift in high fields at $T\simeq 2$~K at different hole doping levels $p$, compared to the spin part of the Knight shift at 285~K.}
 \label{suppl_residual}
\end{figure}

\subsection*{\bf Knight shift determination}

The CDW transition produces a line splitting that has a quadrupole component for both O(2) and O(3) sites (those in bonds oriented parallel to the crystallographic $a$ and $b$ axes, respectively). However, for reasons that are not yet understood, the line splitting has a significant Knight shift contribution only at O(2) sites: the Knight shift contribution to the splitting at O(3) sites is either small or, in the case of the $p=0.109$ sample even totally absent~\cite{Wu15}. Thus, O(3) spectra have little or no asymmetry between high and low frequency satellites, which makes the determination of the Knight shift values more accurate. As far as the Knight shift is concerned, O(3) data are more precise but O(2) data are those sensitive to the CDW. Values in Fig.~2 are for the O(3) sites. Fig.~S3 shows that the field dependence of $K$ at O(2) sites is identical to that of O(3), within error bars. O(2) data are lower than those for O(3), which we attribute to a systematic error in the fitting, due to the uncertainty generated by the absence of splitting of the low-frequency satellites.

It should be noted that electronic bound states in high fields make the distribution of $K$ values asymmetric~\cite{Zhou17}. In low field, the vortex lattice also produces an asymmetric distribution of $K$. Therefore, the Knight shift extracted from the positions of maximum intensity corresponds, strictly speaking, to the mode of the skewed distribution rather than to its mean value. The difference between mean and mode values is however small compared to our error bars so that the extracted $K$ values should be considered as spatially averaged, mean, Knight shifts.

\subsection*{\bf Orbital shift subtraction}

$K_{\rm spin}$ values are obtained by subtracting the so-called orbital part $K_{\rm orb}$ and the diamagnetic shielding contribution $K_{\rm dia}$ from the measured total $K$:
\begin{equation}
K = K_{\rm spin} + K_{\rm orb} + K_{\rm dia}
\end{equation}
$K_{\rm dia}$ should be identical at different $^{17}$O positions. Therefore, the absence of any temperature dependence of $K$ at the apical O(4) sites (Fig.~S1) indicates that $K^{\rm dia}$ is negligible at the fields used here.

Accurately determining $K_{\rm orb}$, on the other hand, is difficult. Being considered to be dominated by Van-Vleck paramagnetism and to be $T$ independent in the cuprates, $K_{\rm orb}$ is usually taken as the value of $K$ in the limit $T\rightarrow 0$. However, because of nodal excitations, this is only true in the limit $H\rightarrow 0$ and given the weakness of the $^{17}$O signal at low fields in the superconducting state, $K_{\rm orb}$ is not directly measurable. Therefore, we used the most reliable values in the literature: $^{17}K^{c}_{\rm orb}=-0.014$\% and $^{17}K^{b}_{\rm orb}=0.044$\%, as determined in ref.~\cite{Takigawa89}, together with the relation $K_{\rm orb}(\theta)=K^c_{\rm orb} \cos^2(\theta)+K^b_{\rm orb} \sin^2(\theta)$.

\subsection{\bf Constraints on $K^{\rm orb}$ values}

The $K_{\rm orb}$ values used in the analysis are consistent with constraints imposed by the measured anisotropy of the Knight shift above $T_c$. The analysis is as follows. The principal component of the measured total shift $K$ along an axis $i$ ($i$ = $a$, $b$, $c$) is expressed as:
\begin{equation}
\label{f1}
K_{i}\left( T \right)={{A}_{i}}\chi (T)+K_{i}^{\text{orb}}
\end{equation}
This then gives the two following relations for the anisotropy between the $c$ axis and each the two planar directions:
\begin{equation}
\label{f1}
\begin{aligned}
  & K_{c}={{{A}_{c}}}/{{{A}_{a}}}\;\cdot K_{a}+\left( K_{c}^{\text{orb}}-{{{A}_{c}}}/{{{A}_{a}}}\;\cdot K_{a}^{\text{orb}} \right) \\
 & K_{c}={{{A}_{c}}}/{{{A}_{b}}}\;\cdot K_{b}+\left( K_{c}^{\text{orb}}-{{{A}_{c}}}/{{{A}_{b}}}\;\cdot K_{b}^{\text{orb}} \right) \\
\end{aligned}
\end{equation}

In Fig. S3, we scaled the O(3) Knight shift measured along the $a$, $b$ and $c$ axes of ortho-II YBa$_2$Cu$_3$O$_{6.56}$, as shown in the legend. From this, we deduce the following anisotropy ratios of hyperfine coupling constants: $A_c$/$A_a$ = 1.07 and $A_c$/$A_b$ = 0.65.

These values are in good agreement with values given in ref.~\cite{Yoshinari90}. 
More importantly, the scaling also imposes that the orbital shift anisotropy is such that $K_{c}^{\text{orb}}-1.07K_{a}^{\text{orb}}=-0.02\%$ and $K_{c}^{\text{orb}}-0.65K_{b}^{\text{orb}}=-0.039\%$. These values are again totally consistent with the most precise $K^{\rm orb}$ values in the literature (ref.~\cite{Takigawa89}).

\end{document}